\documentclass[twoside]{dis09}
\usepackage[latin1]{inputenc}
\usepackage[dvips]{graphicx,epsfig,color}
\usepackage{wrapfig,rotating}
\usepackage{amssymb,amsmath,array}

\begin{document}
\title{Timelike Compton Scattering in Ultraperipheral Collisions}
\author{B. Pire$^1$, L. Szymanowski $^2$, J. Wagner $^2$
\thanks{Presented by J.Wagner at the DIS 2009 Conference, 26-30 April 2009, Madrid \cite{url_Wagner}.}
%
\vspace{.3cm}\\
%
1- {\'E}cole Polytechnique, CNRS - CPHT \\
 91128 Palaiseau - France
%
\vspace{.1cm}\\
2- Soltan Institute for Nuclear Studies - Theoretical Physics Department \\
Ho\.{z}a 69, 00-681 Warsaw - Poland\\
}

\maketitle

\begin{abstract}
Exclusive photoproduction of dileptons, $\gamma N\to
\ell^+\!\ell^- \,N$, is and will be measured in ultraperipheral
collisions at hadron colliders.
We demonstrate that the timelike deeply virtual Compton scattering (TCS)
mechanism  $\gamma q \to \ell^+\!\ell^- q $, where
the lepton pair comes from the subprocess      $\gamma q \to \gamma^* q $,
dominates in some accessible kinematical regions, thus opening a new 
way  to study generalized parton distributions (GPD)    in     the nucleon at
small skewedness. 
\end{abstract}
\section{Introduction.}
Much theoretical and experimental progress  has recently
been witnessed in  the study of deeply virtual Compton scattering (DVCS),
 i.e., $\gamma^* p \to \gamma p$, 
an exclusive reaction where generalized parton
distributions (GPDs) factorize from perturbatively calculable coefficient functions, when
the virtuality of the incoming photon is high enough~\cite{Muller:1994fv}.
It is now recognized that the measurement of GPDs should contribute in a decisive way to
our understanding of how quarks and gluons build 
hadrons~\cite{gpdrev}. In particular the transverse
location of quarks and gluons become experimentally measurable via the transverse momentum dependence of the GPDs \cite{Burk}.  In our work \cite{Pire:2008ea} we study the "inverse" process, 
 $$\gamma(q) N(p) \to \gamma^*(q') N(p') \to l^-(k) l^+(k') N(p')$$
\begin{wrapfigure}{r}{0.5\columnwidth}
\centerline{\includegraphics[width=0.45\columnwidth]{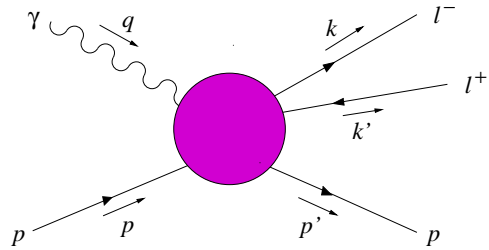}}
\caption{Real photon-proton scattering into a lepton pair and a proton.}
\label{refig}
\end{wrapfigure}
   at small $t = (p'-p)^2$ and large \emph{timelike} virtuality $(k+k')^2=q'^2 = Q'^2$ of the final state
 dilepton, timelike Compton scattering (TCS) \cite{TCS}, which shares many features with DVCS. 


The possibility to use high energy hadron colliders 
as powerful sources of quasi real photons in ultraperipheral collisions has
recently been emphasized \cite{UPC}. This should allow the study of many aspects
of photon proton and photon photon collisions at high energies, already
at the Tevatron and at RHIC but in particular at the
LHC \cite{UPCLHC}    even if the nominal luminosity is not achieved during
its first years of operation. The high luminosity and energies of these
photon beams opens a new kinematical domain for the study of TCS , and
thus to the hope of determining GPDs in the small skewedness ($\xi$) region, which is  complementary  to the determination
 of the large $\xi$ quark  GPDs at lower energy electron accelerators such as
JLab. Moreover, the crossing from 
 a spacelike to a timelike probe is an important test of the understanding of QCD 
 corrections, as shown by the history of the understanding of the Drell-Yan reaction
  in terms of QCD.
  
\section{Photoproduction of a lepton pair}
\begin{figure}[t]
\begin{center}
     \epsfxsize=0.85\textwidth
      \epsffile{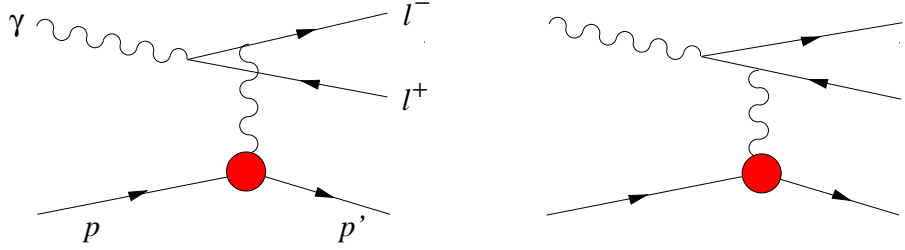}
\caption{The Feynman diagrams for the Bethe-Heitler amplitude.}
\label{bhfig}
\end{center}
\end{figure}
\begin{figure}[b!]
\begin{center}
    \epsfxsize=0.35\textwidth
     \epsffile{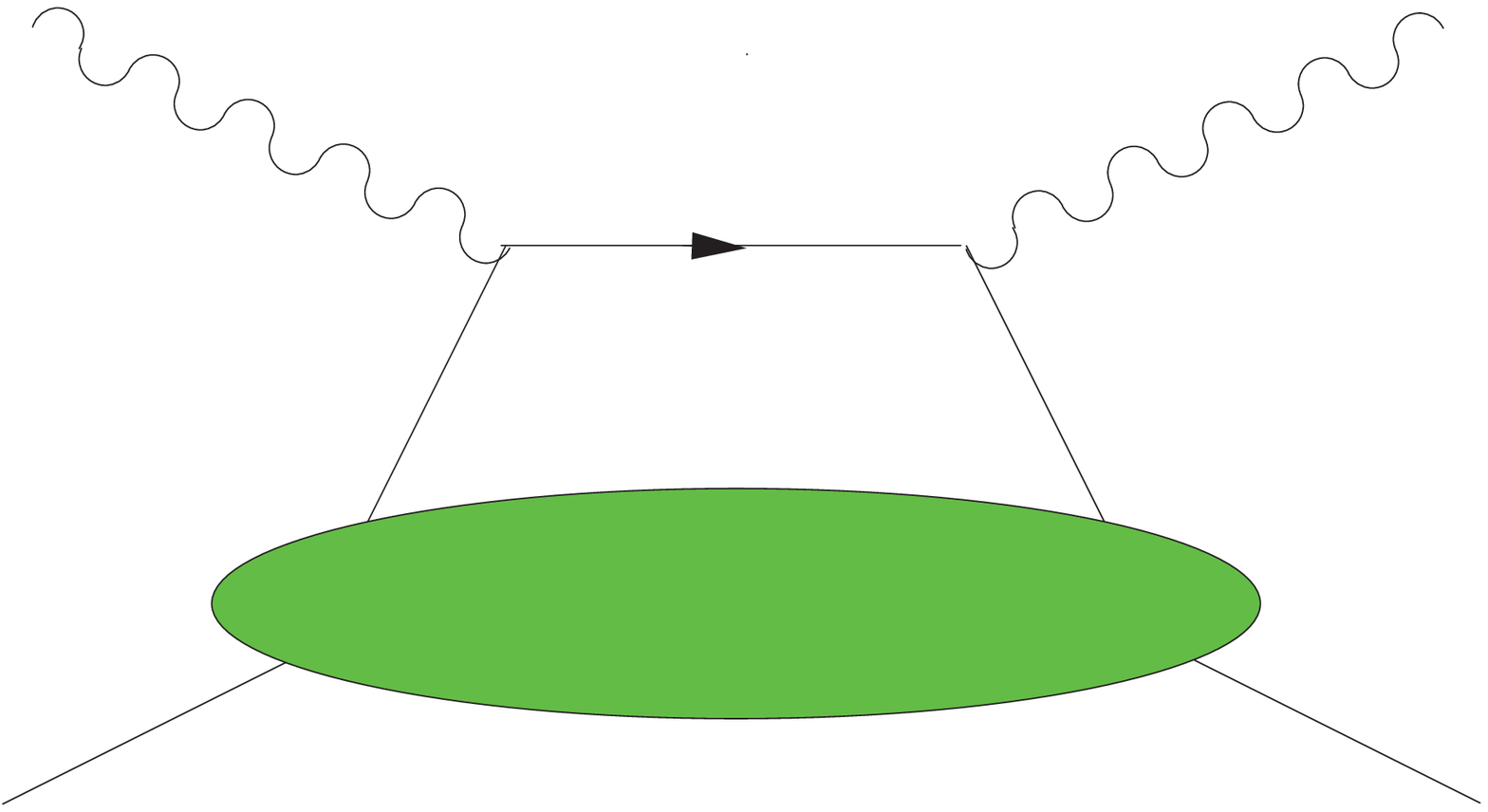}
\hspace{0.05\textwidth}
    \epsfxsize=0.35\textwidth
    \epsffile{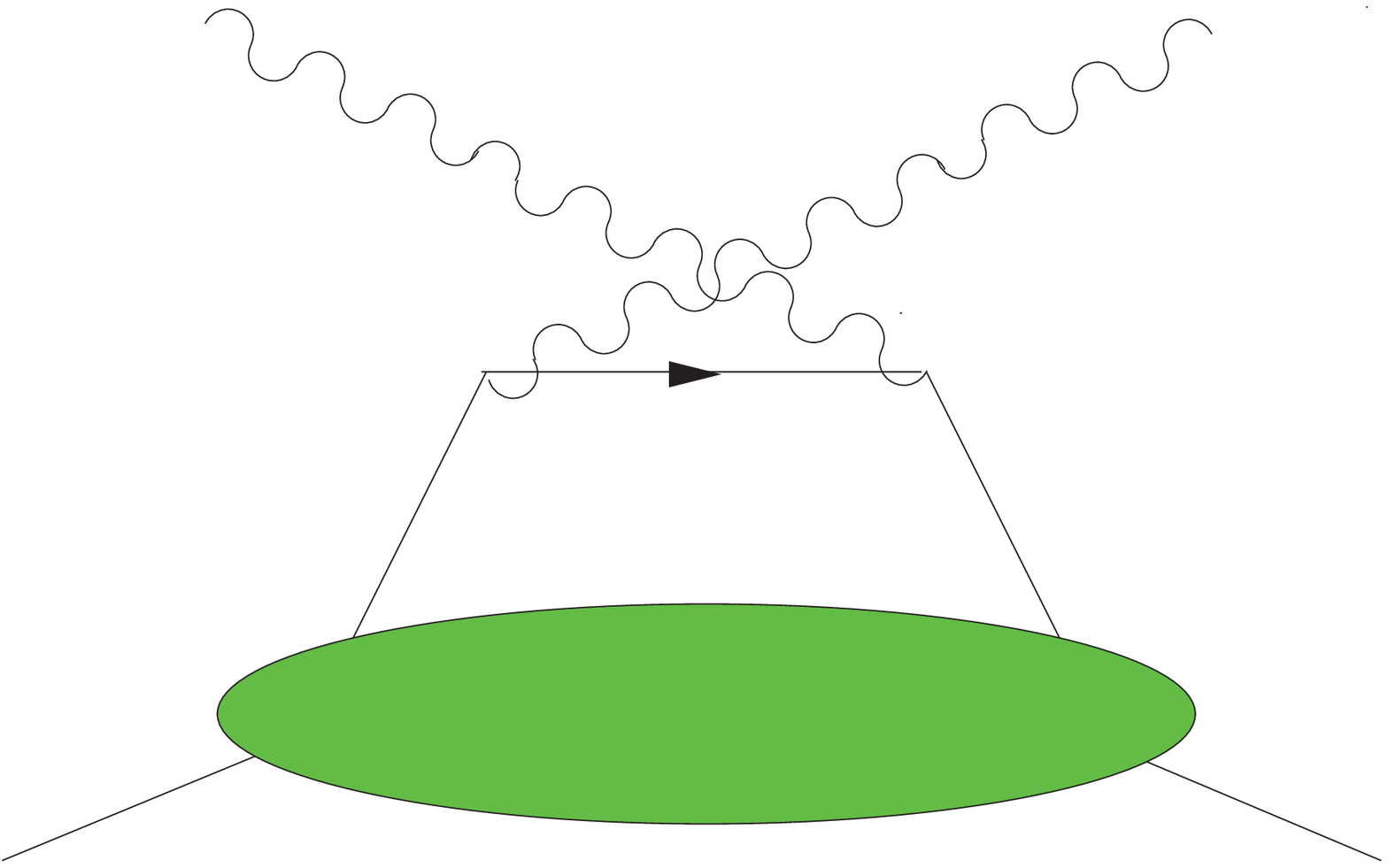}
\caption{Handbag diagrams for the Compton process in the scaling limit.}
\label{haba}
\end{center}
\end{figure}
The physical process where to observe TCS, is photoproduction of a heavy lepton pair, $\gamma N \to \mu^+\!\mu^-\, N$ or $\gamma N \to e^+\!e^-\, N$, shown in Fig.~\ref{refig}. As in the case of DVCS, the Bethe-Heitler (BH)
mechanism 
contributes at the amplitude level. 
This process has a very peculiar angular dependence and overdominates the TCS process if
one blindly integrates over the final phase space. One may however choose kinematics where 
the amplitudes of the two processes are of the same order of magnitude, and either subtract the 
well-known Bethe-Heitler process or use specific observables sensitive to the interference of the two amplitudes. 
The Bethe-Heitler amplitude is  calculated from the two Feynman
diagrams in Fig.~\ref{bhfig}. 
Neglecting masses and $t$  compared to
terms going with $s$ or $Q^{\prime 2}$, the  Bethe Heitler contribution to the unpolarized
$\gamma p$ cross section is ($M$ is the proton mass) 
\begin{eqnarray}
\label{approx-BH}
\frac{d \sigma_{BH}}{d {Q^\prime}^2 d t d \cos \theta} 
&\approx & 2 \alpha^3 \frac{1}{-t {Q'}^4} \frac{1+\cos ^2 \theta}{1-\cos ^2\theta} 
\left(F_1(t)^2 - \frac{t}{4M_p^2} F_2(t)^2\right) ,
\end{eqnarray}
provided we stay away from the kinematical region where the  product  of lepton propagators goes 
to zero at very small $\theta$ ($F_1(t)$ and $F_2(t)$ are Dirac and Pauli nucleon form factors). The interesting physics program thus imposes a
cut on $\theta$ to stay away from the region where the Bethe Heitler  cross section becomes
extremely large.

\begin{figure}[t!]
\begin{center}
\epsfxsize=0.4\textwidth
\epsffile{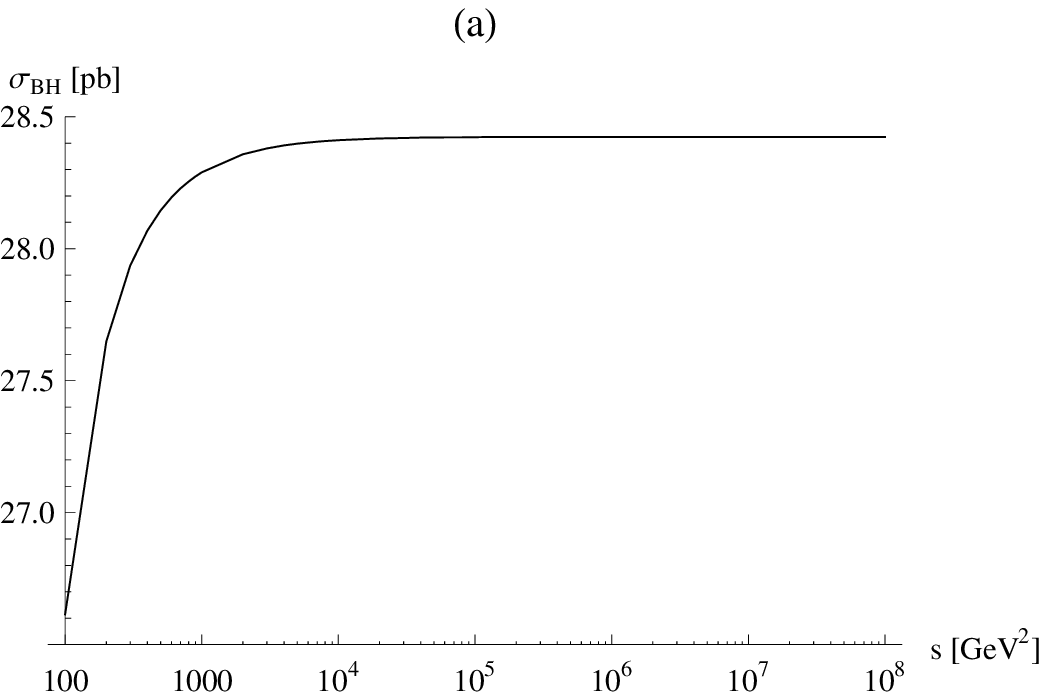}
\hspace{0.05\textwidth}
\epsfxsize=0.39\textwidth
\epsffile{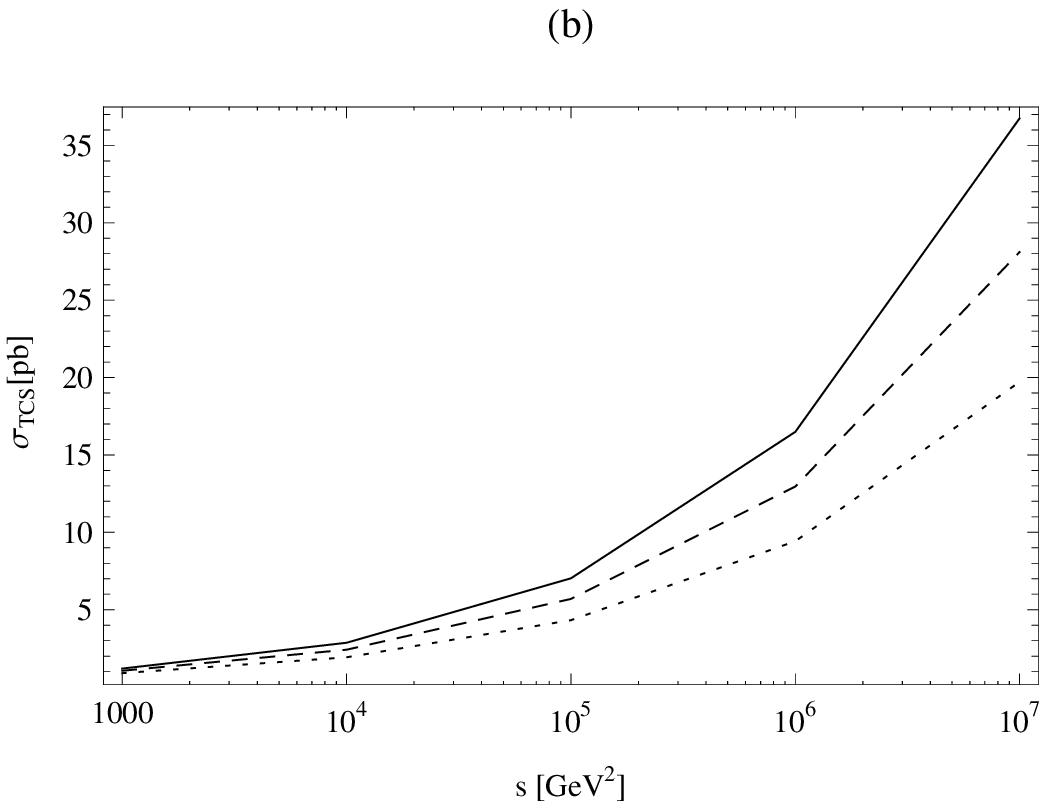}
\caption{ a) The BH cross section integrated over $\theta \in [\pi/4,3
\pi/4]$, $\varphi \in [0, 2\pi]$ , $Q'^2 \in [4.5,5.5]\,{\rm GeV}^2$, $|t| \in [0.05,0.25] \,{\rm GeV}^2$, as a function of $\gamma p$ c.m. energy squared $s$.
b) $\sigma_{TCS}$ as a function of $\gamma p$ c.m. energy squared $s$, for GPD parametrization based on the 
GRVGJR2008 NLO PDF, for different factorization scales $\mu_F^2 = 4$ (dotted), $5$ (dashed), $6$ (solid) $\,{\rm GeV}^2$.
} 
\label{BHs}
\end{center}
\end{figure}
In the region where the final photon virtuality is large, the Compton amplitude is given by the convolution of hard scattering coefficients, calculable in
perturbation theory, and generalized parton distributions, which describe the nonperturbative physics of the process. To leading order
in $\alpha_s$ one then has the dominance of  the quark handbag diagrams of Fig.~\ref{haba}.  
\begin{eqnarray}
\frac{d \sigma_{TCS}}{d {Q^\prime}^2 d \Omega d t} 
\approx \frac{\alpha^3}{8 \pi} \frac{1}{s^2} \frac{1}{{Q'}^2}
\left(\frac{1+\cos^2\theta}{4}\right)
2(1-\eta^2) \left(|{\cal H}|^2+|\tilde{\cal H}|^2\right) , 
\label{eq:Capprox}
\end{eqnarray}
where $\cal{H}$ and $\tilde{\cal H}$ are Compton formfactors, defined as in \cite{TCS}, and $\eta$ is the skewedness parameter related to the Bjorken variable $\tau = Q'^2/s$ by $\eta= \tau/(2-\tau)$.
Full BH and TCS cross section as a functions of c.m. energy squared $s$ are shown on Fig. \ref{BHs}. 
Since the amplitudes for the Compton and Bethe-Heitler
processes transform with opposite signs under reversal of the lepton
charge,  
it is possible to project out
the interference term through a clever use of
 the angular distribution of the lepton pair. 
The interference part of the cross-section for $\gamma p\to \ell^+\ell^-\, p$ with 
unpolarized protons and photons is given at leading order by
\begin{eqnarray}
   \label{intres}
\frac{d \sigma_{INT}}{dQ'^2\, dt\, d\cos\theta\, d\varphi}
= {}-
\frac{\alpha^3_{em}}{4\pi s^2}\, \frac{1}{-t}\, \frac{M}{Q'}\,
\frac{1}{\tau \sqrt{1-\tau}}\,
  \cos\varphi \frac{1+\cos^2\theta}{\sin\theta}
     {\mathrm{Re}\,}\tilde{M}^{--} \; ,
\end{eqnarray}
with ($-t_0 = 4\eta^2 M^2 /(1-\eta^2)$):
\begin{equation}
\label{mmimi}
\tilde{M}^{--} = \frac{2\sqrt{t_0-t}}{M}\, \frac{1-\eta}{1+\eta}\,
\left[ F_1 {\cal H}_1 - \eta (F_1+F_2)\, \tilde{\cal H}_1 -
\frac{t}{4M^2} \, F_2\, {\cal E}_1 \,\right].
\end{equation}

\begin{figure}[t]
\begin{center}
\epsfxsize=0.45\textwidth
\epsffile{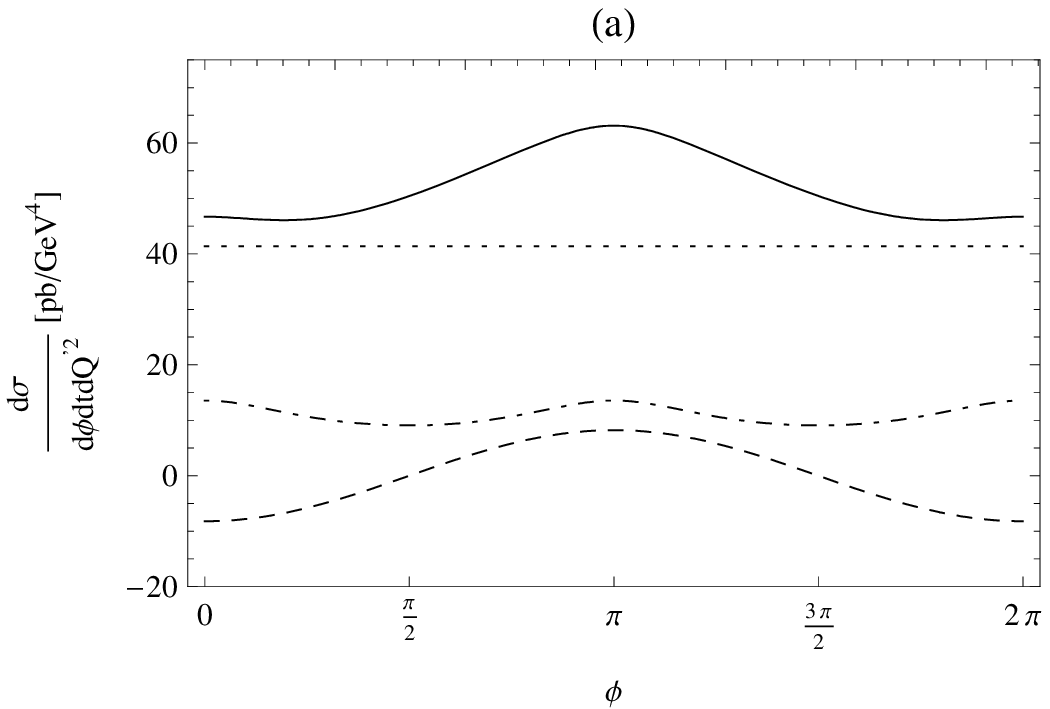}
\hspace{0.05\textwidth}
\epsfxsize=0.45\textwidth
\epsffile{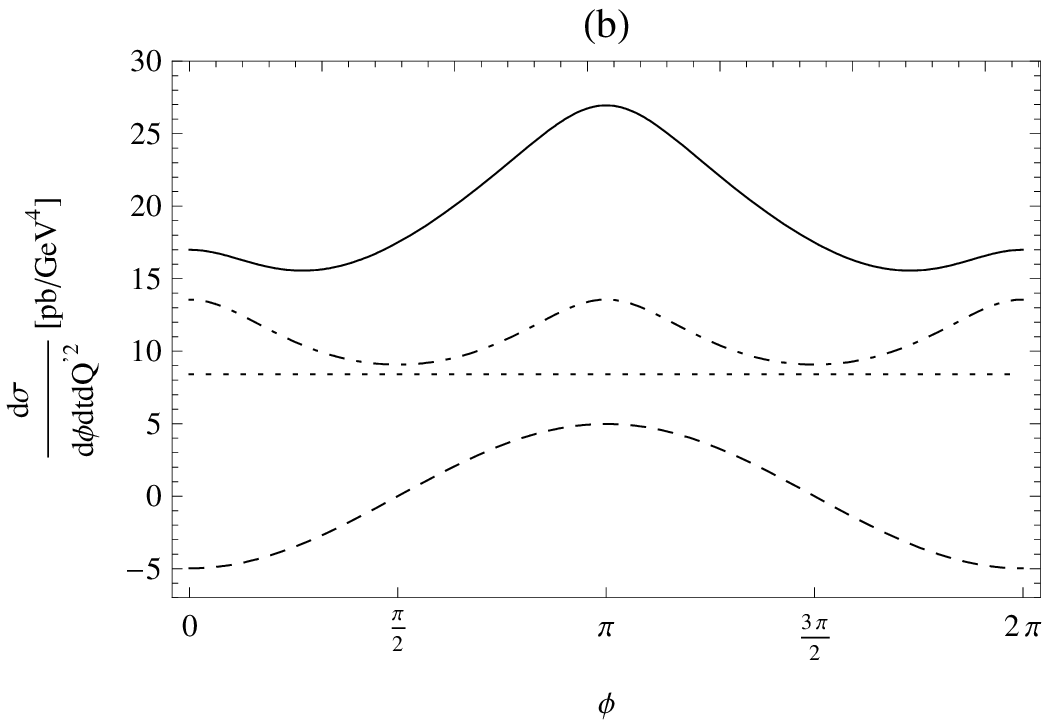}
\vspace{0.05\textwidth}
\epsfxsize=0.75\textwidth
\epsffile{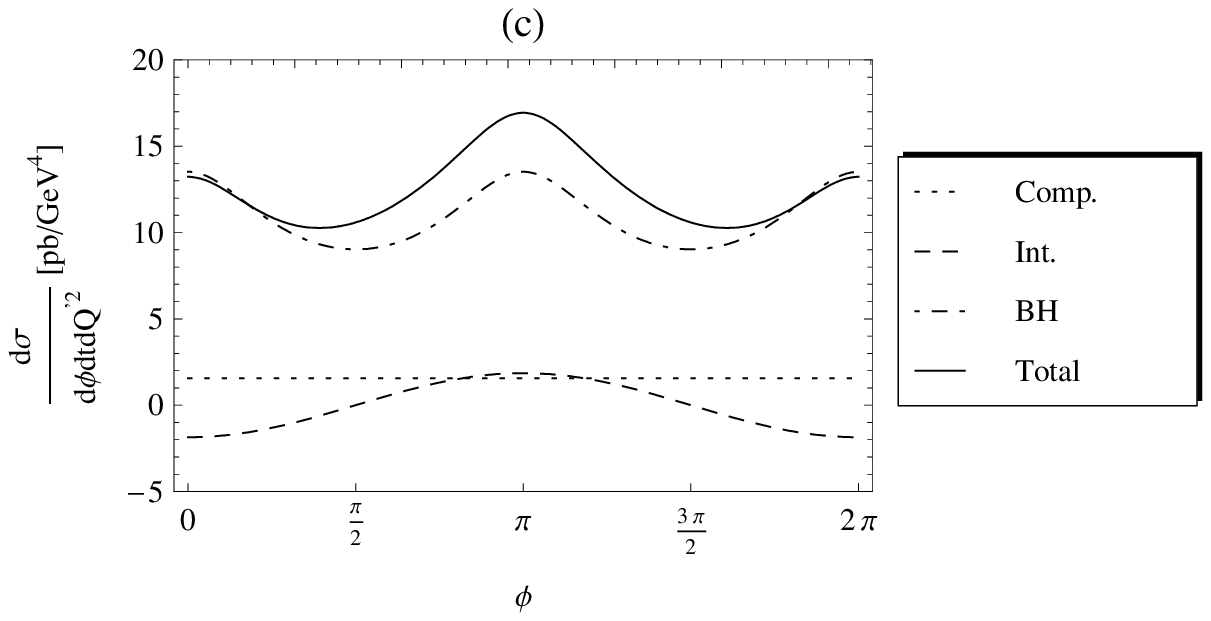}
\caption{
The differential cross sections (solid lines) for $t =-0.2 \,{\rm GeV}^2$, ${Q'}^2 =5 \,{\rm GeV}^2$ and integrated 
over $\theta = [\pi/4,3\pi/4]$, as a function of $\varphi$, for $s=10^7 \,{\rm GeV}^2$ (a), 
$s=10^5 \,{\rm GeV}^2$(b), $s=10^3 \,{\rm GeV}^2$ (c) with $\mu_F^2 = 5 \,{\rm GeV}^2$. We also display  the
Compton (dotted), Bethe-Heitler (dash-dotted) and Interference (dashed) contributions. 
}
\label{Interf}
\end{center}
\end{figure}

Figure \ref{Interf} shows the interference contribution to the cross section in comparison to the Bethe Heitler and Compton processes, for various values of c.m. energy squared $s = 10^7 \,{\rm GeV}^2,10^5 \,{\rm GeV}^2,10^3 \,{\rm GeV}^2$. We observe that for large energies the Compton process dominates, whereas for $s=10^5 \,{\rm GeV}^2$ all contributions are comparable.
\section{Full cross section}
The cross section for photoproduction in hadron collisions is given by:
\begin{equation}
\sigma_{pp}= 2 \int \frac{dn(k)}{dk} \sigma_{\gamma p}(k)dk
\end{equation}
where $\sigma_{\gamma p} (k)$ is the cross section for the 
$\gamma p \to pl^+l^-$ process and $k$ is the photon energy. 
$\frac{dn(k)}{dk}$ is an equivalent photon flux. 
The relationship between $\gamma p$  energy squared $s$ and k is given by $
s \approx 2\sqrt{s_{pp}}k \nonumber$, 
where $s_{pp}$ is the proton-proton  energy squared ($\sqrt{s_{pp}} = 14 {\,{\rm TeV}}$)

The Bethe - Heitler contribution to $\sigma_{p p}$, integrated over  $\theta = [\pi/4,3\pi/4]$, $\phi = [0,2\pi]$, $t =[-0.05 \,{\rm GeV}^2,-0.25 \,{\rm GeV}^2]$, ${Q'}^2 =[4.5 \,{\rm GeV}^2,5.5 \,{\rm GeV}^2]$, and photon energies $k =[20,900]\,{\rm GeV} $  gives:
\begin{equation}
\sigma_{pp}^{BH} = 2.9 {\,{\rm pb}} \;.
\end{equation}  
The Compton contribution (calculated with NLO GRVGJR2008 PDFs, and $\mu_F^2 = 5 \,{\rm GeV}^2$) gives:
\begin{equation}
\sigma_{pp}^{TCS} = 1.9 {\,{\rm pb}}\;. 
\end{equation}
We have choosen the range of photon energies in accordance with expected capabilities to tag photon energies
at the LHC. This amounts to a large rate of order of $10^5$ events/year at the LHC with its nominal 
luminosity ($10^{34}\,$cm$^{-2}$s$^{-1}$). 
\section{Conclusions}
Timelike Compton scattering in ultraperipheral collisions at hadron colliders opens a new way to measure generalized parton distributions. We have found sizeable rates of events at LHC, even for the lower luminosity which can be achieved in the first months of run. Our work has to be supplemented by studies of higher order contributions which  will involve the gluon GPDs.
\vspace{0.5 cm}

\section*{Acknowledgements} This work is partly supported by the ECO-NET program, contract 
18853PJ, the French-Polish scientific agreement Polonium and the Polish Grant N202 249235.

\begin{footnotesize}


%

\end{footnotesize}


\end{document}